\documentclass[%
    reprint,
superscriptaddress,
 amsmath,amssymb,
prb,
]{revtex4-2}

\usepackage{graphicx}
\usepackage{dcolumn}
\usepackage{bm}
\usepackage{siunitx}

\begin{document}

\title{Identifiying the domain wall spin structure in current-induced switching of antiferromagnetic NiO/Pt}

\author{C. Schmitt}
 \email{Christin.Schmitt@Uni-Mainz.de}
 \affiliation{Institute of Physics, Johannes Gutenberg-University Mainz, 55099 Mainz, Germany}

 \author{L. Sanchez-Tejerina}
 \affiliation{Department of Mathematical and Computer Sciences, Physical Sciences and Earth Sciences, University of Messina, 98166 Messina, Italy}
  \affiliation{Present address: Grupo de Investigación en Aplicaciones del Láser y Fotónica, Departamento de Física Aplicada, Universidad de Salamanca, E-37008, Salamanca, Spain}
\author{M. Filianina}
 \affiliation{Institute of Physics, Johannes Gutenberg-University Mainz, 55099 Mainz, Germany}
 \affiliation{Graduate School of Excellence Materials Science in Mainz, 55128 Mainz, Germany}
 \affiliation{Present address: Department of Physics, AlbaNova University Center, Stockholm University, S-106 91 Stockholm, Sweden}
\author{F. Fuhrmann}
 \affiliation{Institute of Physics, Johannes Gutenberg-University Mainz, 55099 Mainz, Germany}
\author{H. Meer}
 \affiliation{Institute of Physics, Johannes Gutenberg-University Mainz, 55099 Mainz, Germany}
\author{R. Ramos}
 \affiliation{WPI-Advanced Institute for Materials Research, Tohoku University, Sendai 980-8577, Japan}
 \affiliation{Present address: Center for Research in Biological Chemistry and Molecular Materials (CIQUS), Departamento de Química-Física, Universidade de Santiago de Compostela, Santiago de Compostela 15782, Spain}
 
\author{F. Maccherozzi}
 \affiliation{Diamond Light Source, Harwell Science and Innovation Campus, Didcot OX11 0DE, United Kingdom}%
 
\author{D. Backes}
 \affiliation{Diamond Light Source, Harwell Science and Innovation Campus, Didcot OX11 0DE, United Kingdom}%
 
\author{E. Saitoh}
\affiliation{WPI-Advanced Institute for Materials Research, Tohoku University, Sendai 980-8577, Japan}
 \affiliation{Institute for Materials Research, Tohoku University, Sendai 980-8577, Japan}
 \affiliation{The Institute of AI and Beyond, The University of Tokyo, Tokyo 113-8656, Japan}
 \affiliation{Center for Spintronics Research Network, Tohoku University, Sendai 980-8577, Japan}
 \affiliation{Department of Applied Physics, The University of Tokyo, Tokyo 113-8656, Japan}

\author{G. Finocchio}
\affiliation{Department of Mathematical and Computer Sciences, Physical Sciences and Earth Sciences, University of Messina, 98166 Messina, Italy}

\author{L. Baldrati}
 \affiliation{Institute of Physics, Johannes Gutenberg-University Mainz, 55099 Mainz, Germany}
\author{M. Kläui}
\email{Klaeui@Uni-Mainz.de}

 \affiliation{Institute of Physics, Johannes Gutenberg-University Mainz, 55099 Mainz, Germany}
 \affiliation{Graduate School of Excellence Materials Science in Mainz, 55128 Mainz, Germany}

\date{\today}

\begin{abstract}
The understanding of antiferromagnetic domain walls, which are the interface between domains with different Néel order orientations, is a crucial aspect to enable the use of antiferromagnetic materials as active elements in future spintronic devices. In this work, we demonstrate that in antiferromagnetic NiO/Pt bilayers circular domain structures can be generated by switching driven by electrical current pulses. The generated domains are T-domains, separated from each other by a domain wall whose spins are pointing toward the average direction of the two T-domains rather than the common axis of the two planes. Interestingly, this direction is the same for the whole circular domain indicating the absence of strong Lifshitz invariants. The domain wall can be micromagnetically modeled by strain distributions in the NiO thin film induced by the MgO substrate, deviating from the bulk anisotropy. From our measurements we determine the domain wall width to have a full width at half maximum of $\Delta = 98 \pm \SI{10}{nm}$, demonstrating strong confinement. 
\end{abstract}

\maketitle

\newpage
\section{\label{sec:intro}Introduction\protect}

Spintronic devices to date mostly rely on ferromagnets (FMs) as active elements to store magnetic information. However, antiferromagnets (AFMs) possess several advantages over FMs for use in applications such as spin dynamics in the THz regime, high stability in the presence of external magnetic fields and a lack of magnetic stray fields. Therefore, AFMs are prime candidates to replace FMs  as active elements in future spintronic devices \cite{Baltz2018, kampfrath2011coherent}. In particular insulating antiferromagnetic materials are promising candidates for the development of low power devices, because their low damping allows for
the transport of pure spin currents over long distances \cite{lebrun2020long}.\\
However, a key requirement in view of applications is efficient and reliable reading and writing of magnetic information encoded in the orientation of the antiferromagnetic Néel order \textbf{n}. It has been established, that electrical current pulses through an adjacent heavy metal layer can induce a reorientation of the antiferromagnetic order in insulating AFMs \cite{Moriyama2018, chen2018antidamping, baldrati2019mechanism}. However, the switching mechanism is highly debated \cite{Moriyama2018, chen2018antidamping, baldrati2019mechanism, zhang2019quantitative, Meer}: On the one hand, for insulating AFMs with strong magnetostriction, such as NiO \cite{aytan2017spin}, the writing of the Néel vector is reported to be dominated by a thermomagnetoelastic switching mechanism \cite{zhang2019quantitative, Baldrati2020, Meer} and especially for high current densities electromigration effects in the heavy metal layer were observed \cite{churikova2020non}. On the other hand, switching mechanisms based on spin-current-induced domain wall motion are also theoretically predicted for this class of materials \cite{shiino2016antiferromagnetic, sanchez2020dynamics, baldrati2018full} and switching experiments with current densities smaller than the ones necessary for electromigration were performed \cite{arpaci2021observation}.\\
The domain walls (DWs) in antiferromagnets provide a key to understanding the magnetic microstructure \cite{hubert2008magnetic} and the spin structure influences the thermal stability \cite{shpyrko2007direct}, magnetoresistance \cite{jaramillo2007microscopic, wornle2019current} and exchange bias when coupled to a FM \cite{radu2008exchange} in the AFMs. Furthermore, the type of DW, Bloch or Néel or a mixture of both \cite{wornle2021coexistence}, affects their response to current-induced spin-torques \cite{hals2011phenomenology, gomonay2016high, shiino2016antiferromagnetic, baldrati2019mechanism, sanchez2020dynamics} and also is crucial to understand the magnon coupling in AFMs \cite{bossini2021ultrafast, gomonay2021linear}.\\
While DWs in FMs are extensively studied \cite{hubert2008magnetic}, the structure of DWs in AFMs is often unknown. Although, the domains have been imaged for various materials \cite{grzybowski2017imaging, sapozhnik2018direct, xu2019imaging, xu2020optical, schreiber2020concurrent, cheong2020seeing}, these studies often do not include the internal DW spin structure. The lack of experimental results is partially due to an insufficient spatial resolution of many techniques. Yet, only a few studies investigate antiferromagnetic domain walls, for example in synthetic antiferromagnets \cite{yang2015domain}, monolayer-thin films \cite{bode2006atomic}, bulk systems of $\mathrm{Cr}_2\mathrm{O}_3$ \cite{wornle2019current}, and NiO \cite{weber2003magnetostrictive, slack1960crystallography, arai2012three} and thin films of CuMnAs \cite{krizek2022atomically}. Theoretically it has been predicted that DWs in thin antiferromagnetic films can become chiral due to Lifshitz invariants \cite{akanda2020interfacial}. Due to the lack of experimental results on antiferromagnetic DWs, this has not been checked experimentally for the predicted systems, such as NiO. However, the internal domain wall structure affects the response of antiferromagnets to excitations and governs the dynamics. This is of critical importance when considering how to read, transfer and write magnetic information in AFMs, and thus revealing the AFM DW spin structure is a key open question.\\
An important material in the context of AFM spintronics is NiO, for which in thin film form the domain wall structure is not known. NiO is considered as a promising material for active elements in spintronic applications, due to the possibility of electrical control and readout of the AFM order \cite{Moriyama2018, hoogeboom2017negative, baldrati2018full, fischer2018spin} and observation of ultrafast currents in the THz regime in NiO/Pt bilayers \cite{kampfrath2011coherent, moriyama2020tailoring}. Bulk NiO shows a simple cubic crystallographic structure above the Néel temperature of  $T_\mathrm{N}= \SI{523}{K}$ \cite{roth1960neutron} but it contracts along the $\langle111\rangle$  directions below $T_\mathrm{N}$. As a consequence the spins are confined into one of the four equivalent ferromagnetic $\{111\}$ planes, which are coupled antiferromagnetically (T-domains). Within each of these T-domains, there are three easy axes along the $\langle112\rangle$ directions (S-domains) leading to 12 possible domain orientations in bulk NiO \cite{nussle2019coupling, roth1960antiferromagnetic, slack1960crystallography}. Therefore, we can distinguish between S-domain walls, where the spin rotates but the crystallographic structure is kept, and T-domain walls, in which the direction of the distortion of the original cube changes from one $[111]$ diagonal to another \cite{roth1960neutron}. In the case of T-domains, the distortion of the two adjacent domains must match, imposing certain restrictions \cite{slack1960crystallography}. Therefore, T-domain walls follow particular crystallographic directions. An understanding of the antiferromagnetic DW configuration in NiO thin films and determining the DW chirality is thus of importance when considering how NiO can be used as an active element in spintronic devices.\\
In this work, we determine the Néel vector orientation by photoemission electron microscopy (PEEM) exploiting the x-ray magnetic linear dichroism (XMLD) effect \cite{Schmitt2021} in epitaxial antiferromagnetic NiO thin films grown epitaxially on MgO(001). Our results demonstrate the possibility to create a circular shaped antiferromagnetic domain by current-induced switching which, therefore, do not comply with the aforementioned crystallographic constrictions for the bulk case. We show by micromagnetic simulations that the non-chiral antiferromagnetic domains that we see in our samples can be explained by an anisotropy originating from the strain distribution present in the thin films. This anisotropy also agrees with the  XMLD images for the DW configuration for which the Néel vector points out of the plane and we quantify the DW width to a full width at half maximum (FWHM) of $\Delta = 98 \pm \SI{10}{nm}$, indicating narrow domain walls and the importance of anisotropy in our thin films.

\section{\label{sec:results}Results\protect}

\begin{figure}
 \includegraphics[width=9cm]{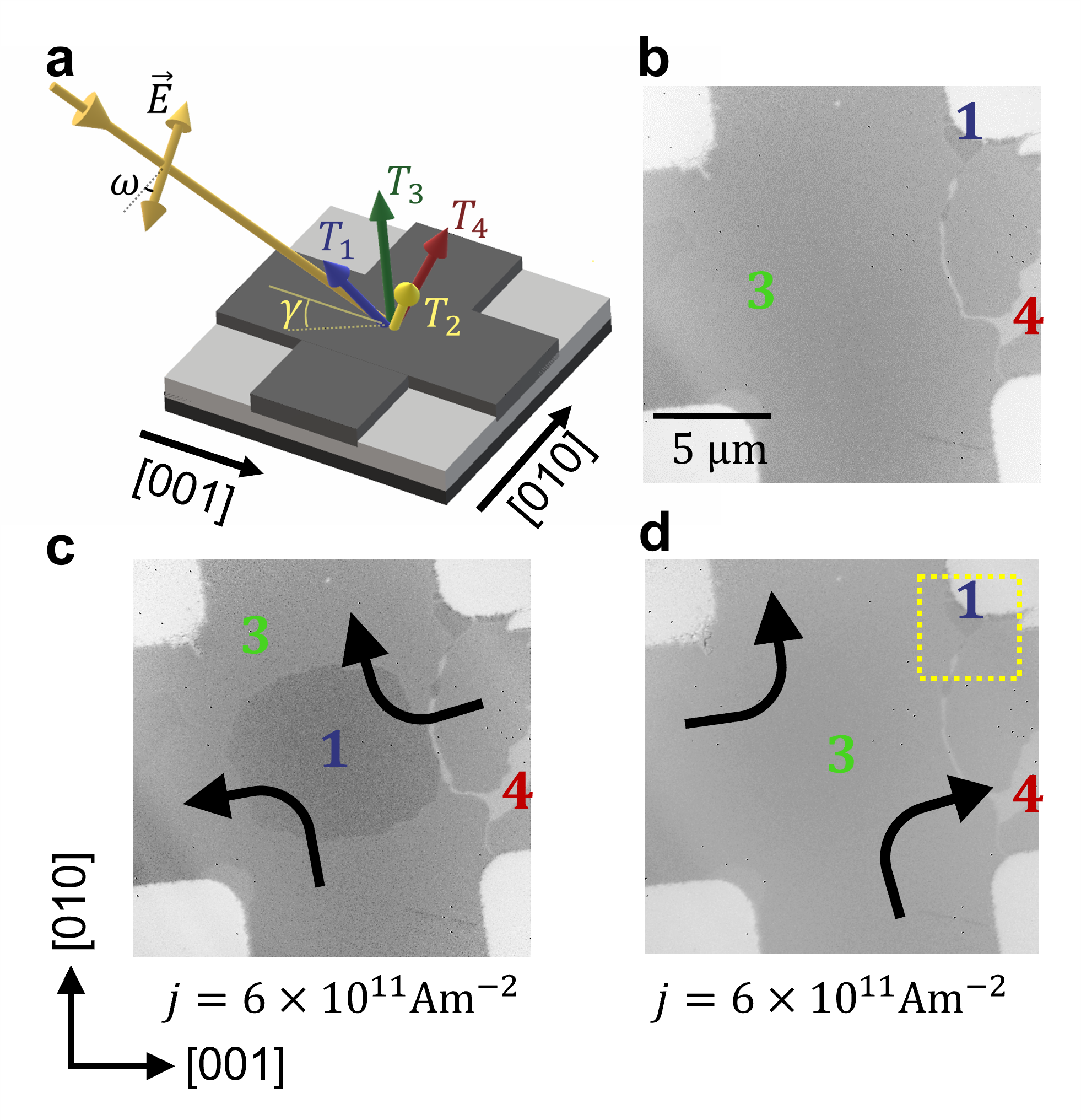}
 \caption{(a) Experimental layout and pulsing scheme. The angles defining the linear polarization vector and the Néel vector are defined with respect to the crystallographic axes. (b) The initial state of the sample before an electric pulse was applied. (c) A circular shaped domain after the application of an x-shaped pulse along the direction indicated by the arrows. (d) After applying an x-shaped pulse with equal current density along the perpendicular direction, the circular shaped domain vanishes and the initial state is restored. The yellow square indicates the area that is magnified in Fig. \ref{fig:two}.}
 \label{fig:one}
\end{figure}

To investigate the DW structure and the effect of electrical current pulses on this structure, we have fabricated epitaxial NiO($\SI{10}{nm}$)/Pt($\SI{2}{nm}$) bilayers on MgO(001) substrates, using the protocol described in Ref. \cite{Schmitt2021} where we verified the antiferromagnetic ordering by the observation of x-ray magnetic linear dichroism (XMLD) and negligible x-ray magnetic circular dichroism (XMCD). In order to be able to apply electrical current pulses, a $\SI{10}{\micro m}$ wide Hall cross device is patterned using Ar ion beam etching. Fig. 1(a) shows the device layout. The Hall cross is oriented along the $[100]$ crystallographic axis. To acquire the XMLD-PEEM images, the Ni $L_2$ edge is used and the contrast is calculated as $\mathrm{XMLD} = \frac{I(E_{\mathrm{low}}) - I(E_{\mathrm{high}})}{I(E_{\mathrm{low}}) + I(E_{\mathrm{high}})}$, where $E_{\mathrm{low}} = \SI{869.7}{eV}$ and $E_{\mathrm{high}} = \SI{871.0}{eV}$. By studying the dependence of the XMLD-PEEM contrast on the azimuthal angle $\gamma$ and the angle of the linear polarization of the x-rays $\omega$, analogous to Ref. \cite{Schmitt2021}, we can identify the domains in this sample to be pointing along the $[\pm5 \pm5\ 19]$ directions, as indicated in Fig. \ref{fig:one} (a). This domain configuration allows for DWs between domains with in-plane Néel vector components differing by 90° and a DW between two T-domains with an in-plane Néel vector component with 180° difference. In order to electrically manipulate the Néel vector orientation, current pulses of $\SI{1}{ms}$ duration were applied to the sample. The initial state is largely single domain as Fig. 1(b) shows, in line with the reports of large domains for high-quality bulk NiO \cite{roth1960antiferromagnetic}, and is set by the application of a pulse with a current density $j = 8.0 \times 10^{11}\ \mathrm{A/m^2}$ along the $[010]$ direction. Only at the patterning edge of the Hall cross small domains are nucleated due to changes of the surface anisotropy at the edge due to the patterning \cite{meer2022strain}. With an x-shaped pulse (represented by the black arrows in Fig. 1(c)) with the current direction in the center of the cross along $[\bar{1}10]$ and a current density of $j = 1.25 \times 10^{12}\ \mathrm{A/m^2}$ electrical switching in the center of the Hall cross is achieved and a circular shaped domain is nucleated. An x-shaped pulse (Fig. \ref{fig:one} (d)) with equal current density but perpendicular direction along $[110]$ (in the center of the cross) switches the magnetic order back to the initial state, as seen in Fig. 1(d).\\

\begin{figure}
 \includegraphics[width=9cm]{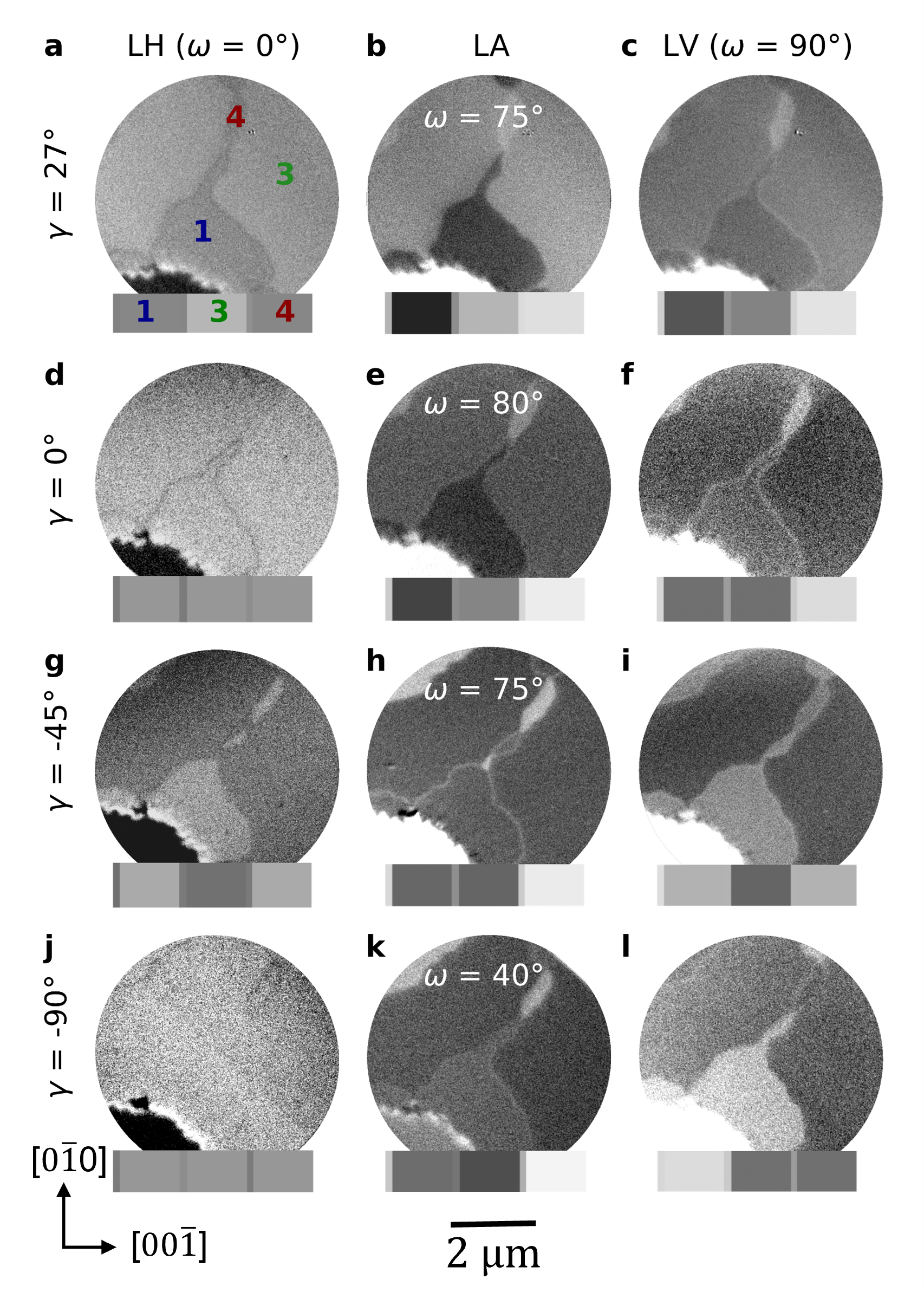}
 \caption{Three level domain contrast and corresponding grey scale simulation for different azimuthal angles $\gamma$ and angles of the beam polarization $\omega$. The DW between two neighboring domains can be seen easily whenever the grey scale of the two domains is similar. The simulated grey scale contrast considering a Néel vector angle with respect the out of plane direction  of $\theta =$ 78.7° $\pm$ 0.3° is depicted below the XMLD-PEEM image. The simulated contrast matches the domain and domain wall contrast for all combinations of azimuthal angle $\gamma$ and beam polarization $\omega$.}
 \label{fig:two}
\end{figure}

Fig. \ref{fig:two} shows a rotated magnification of the domains that are nucleated at the top right corner of the XMLD-PEEM image in Fig. \ref{fig:one} (d) (indicated by a yellow frame). We see three different types of domains nucleated at the edge of the Hall cross for different azimuthal angles $\gamma$ (27°, 0°, -45°, -90°) and three different beam polarizations $\omega$ (0° (LH), 90° (LV) and a linear arbitrary angle (LA)). One sees that when changing $\gamma$ or $\omega$ the grey scale contrast between the three domain types varies along with the projection of the x-ray polarization vector $\mathbf{E}$ on the Néel vector $\mathbf{n}$. The way the XMLD-PEEM signal is calculated leads to the maximum XMLD signal (area with the Néel vector having the largest projection on the \textbf{E}-vector) being represented in white and low XMLD signals (area with the smallest projection of the Néel vector on the \textbf{E}-vector) in black. For certain angles, e.g $\gamma =$ 0°, $\omega =$ 0°, the XMLD-PEEM signal for two, or for all three of the observed domains is equal, meaning that the projection of the different Néel vectors on the x-ray polarization is the same. In this case, the DW in between the domains becomes easily visible (see e.g. Fig. 2 (d)). We can see that the XMLD-PEEM resolution is not good enough to resolve the details of the gradual rotation of the Néel vector. However, from the contrast, we can unambiguously determine the sense of rotation and from a fit of the profile we can extract the domain wall width.\\
Taking into account the data for different $\gamma$ and $\omega$ values, the greyscale contrast between the individual domains and domain walls, angles where the contrast disappears and contrast inversion points, one can obtain information on the rotation direction of the DW. The best agreement between experimental data and contrast simulation is obtained for a canting of the Néel vector of $\theta = $ 78.7° $\pm$ 0.3° out of the sample plane, corresponding to the $[\pm 5\ \pm5\ 33]$ direction. Fig. \ref{fig:two} shows that for this angle, the contrast between domains and domain walls agrees with the simulated contrast for all combinations of azimuthal angles $\gamma$ and polarizations $\omega$. This indicates a rotation of the Néel vector out of the sample plane in comparison to the $[\pm5\ \pm5\ 19]$ orientation of the Néel vector in the magnetic domains \cite{Schmitt2021}. From previous studies on bulk NiO one would expect a rotation of the Néel vector in a way that it crosses the diagonal shared by the two T-domains since in the planes the anisotropy energy density is smallest \cite{weber2003magnetostrictive}. Thus, one could expect the Néel vector to rotate in the \{111\} planes diminishing the anisotropy energy density \cite{gomonay2021linear}. This, however, would lead to a decreased out-of-plane component of the Néel vector $n_z$ within the DW, contrary to the observation here in NiO thin films. The observed increase of the $z$-component in Fig. \ref{fig:two} indicates a rotation of $\mathbf{n}$ along the shortest path from one domain to the other, even though this requires a rotation out of the \{111\} planes, which increases the anisotropy energy density. However, micromagnetic simulations suggest that this configuration, with the rotation of \textbf{n} on the shortest path between the domains, reduces the exchange energy density and the DW width, thus minimizing the total energy. We want to point out that along the entire structures, the DWs show the same contrast, meaning the orientation of the spins in the DWs is the same everywhere in the DW. Therefore, we find that in contrast to predictions \cite{akanda2020interfacial} the DW in our NiO/Pt thin film heterostructure is not chiral. 

Assuming T-domains to be present in the thin films, which show the same anisotropy as in bulk materials, different domains are contracted along different $\langle111\rangle$ directions. This yields a crystallographic constraint for the DW direction. Fig. 3 shows schematically a $T_1$ domain, represented by a cube slightly contracted along the [111] diagonal (indicated by the green line) and a $T_3$ domain, represented by a cube slightly contracted along the $[1\bar{1}1]$ diagonal. These two types of domains can match sharing the (010) plane, forming a twin wall. If the cube is contracted along another [111] diagonal, the two domains cannot share that plane. It follows, when assuming a bulk-like anisotropy that two T-domains cannot form a circular domain. To understand the rotation of the Néel vector within the DW as well as its shape, micromagnetic simulations were performed. As described in the supplementary material \ref{AppendA}, we are taking into account both, the bulk anisotropy as well as the magnetoelastic contributions to the effective field. In this case, it is possible to reproduce the Néel vector direction within the DW but, as described above, this anisotropy imposes strong crystallographic conditions incompatible with the observed circular domain. However, if we neglect the relatively modest bulk anisotropy and consider only the magnetoelastic interaction, it is possible to fit the Néel vector direction within the DW and to remove the crystallographic constraints by considering different shear strains in the two domains.

\begin{figure}
 \includegraphics[width=9cm]{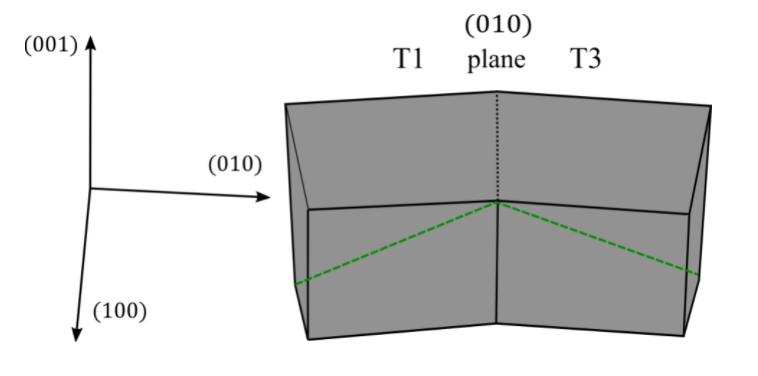}
 \caption{An exaggerated model of the rhombohedral distortion in NiO below the Néel temperature. The green lines along the [111] diagonal and $[1\bar{1}1]$ diagonal indicate the axes along which the $T_1$ and $T_3$ domain are contracted respectively. The two domains can match sharing the (010) plane, forming a twin wall.}
 \label{fig:three}
\end{figure}

\begin{figure*}
 \includegraphics[width=15cm]{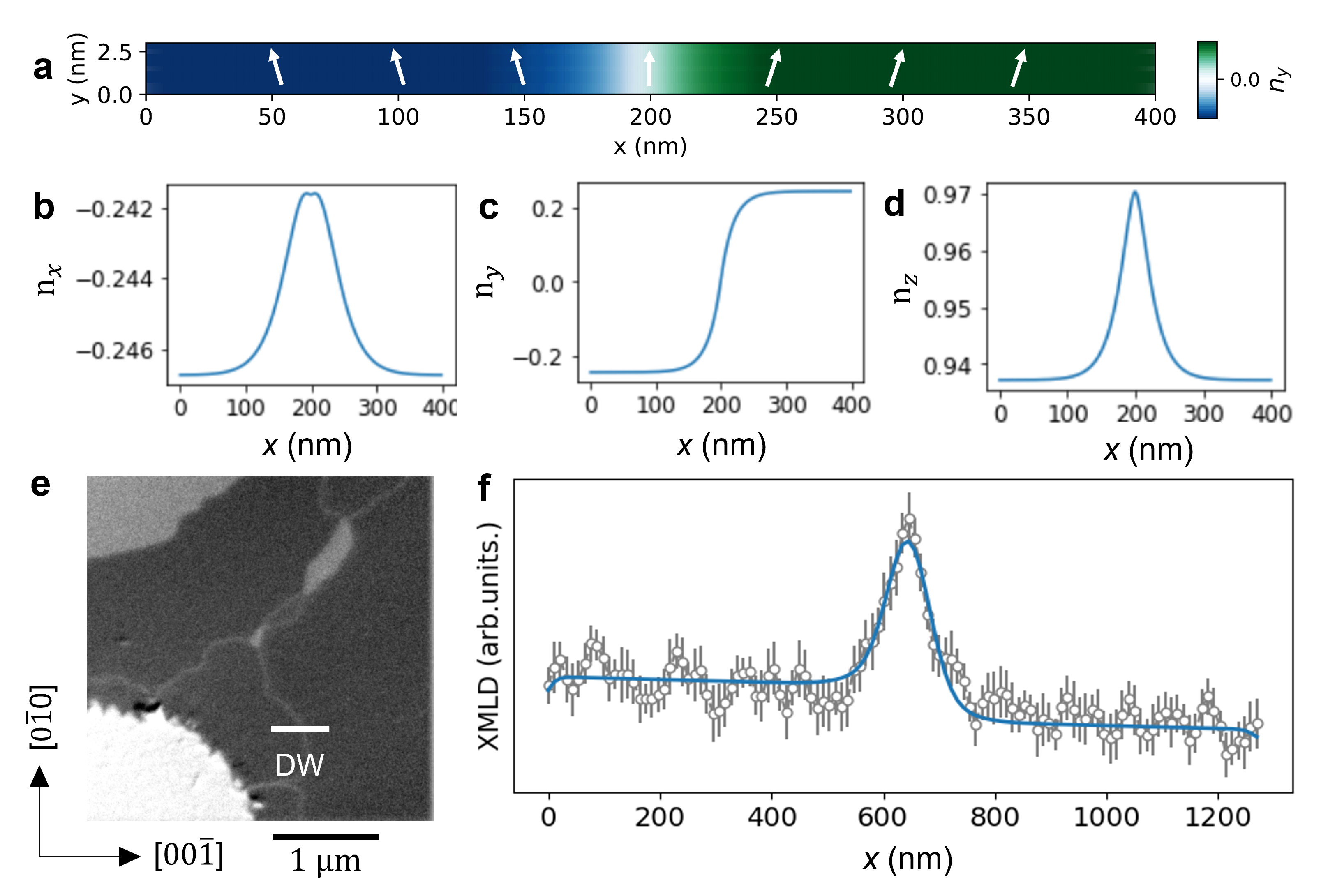}
 \caption{(a) The simulation of a DW within a strip between a blue $T_1$ domain $([\bar{5}\ \bar{5}\ 19])$ and a green $T_3$ domain $([\bar{5}\ 5\ 19])$ allows to read out the line profile of the Néel vector $\mathbf{n}$ change within a DW for (b) $n_x$, (c) $n_y$ and (d) $n_z$. The line profile extracted from the XMLD-PEEM image in (e) can be fitted by a convolution of the theoretical XMLD signal given by the simulation of the DW and the PEEM resolution limit.}
 \label{fig:four}
\end{figure*}

The micromagnetic simulations additionally provide the nanoscale DW profile that agrees with the experimental results, which are of course limited by the finite PEEM resolution. Therefore, a strip consisting of a $T_1$ domain with Néel vector direction along $[\bar{5}\ \bar{5}\ 19]$ at one end and a $T_3$ domain with $\mathbf{n}$ along $[\bar{5}\ 5\ 19]$ at the other end was simulated. The dimension of the strip is $400\times5$ cells with a cell size of $\SI{0.5}{nm}$, resulting in a total simulated volume of $\SI{200}{nm}$ by $\SI{2.5}{nm}$ with a thickness of $\SI{0.5}{nm}$. The result of the simulation, including the DW in between the two domains, is shown in Fig. \ref{fig:four} (a). Here, the white arrows visualise the Néel-vector projection in the \textit{yz}-plane showing an increased out-of-plane component in the center of the DW, as observed experimentally. Fig. \ref{fig:four} (b)-(d) show the extracted Néel vector components as a function of the position within the DW for the three spatial directions, where $x = [100]$, $y = [010]$ and $z = [001]$. The $x$- and $z$-components of the Néel vector along the DW can be approximated by a Gaussian function, while the $y$-component can be represented by a distribution function. The width of the DWs in a magnetic thin film provides significant information about the material and its properties, since the DW width is determined by the ratio of anisotropy and exchange energy constant. However, it needs to be noted that the resolution of the PEEM microscope has an important influence on how wide certain structures appear in the XMLD-PEEM images and this needs to be taken into account when determining the DW width. Therefore, in order to be able to make a statement about the DW width, one must first determine the resolution limit of the microscope in the used configuration as it is shown in the supplementary material \ref{AppendB}. For the determination of the DW width the finite spatial resolution can be taken into account by fitting the signal convoluted with a Gaussian function with FWHM corresponding to the resolution limit \cite{boulle2016room}. For the fit the \textbf{E}-vector is considered constant and given by the XMLD-PEEM image the data was extracted from ($\gamma =$ 45°, $\omega$ = 75°) and $\mathbf{n}$ is defined by the DW profile. A linear function is added in order to depict the intensity gradient in the XMLD-PEEM image which can be observed in Fig. \ref{fig:four} (f) as a linear decay. This yields a FWHM for the DWs between two 90° T-domains of $98 \pm \SI{10}{nm}$. We would like to point out that this determined width is much larger than the resolution of the PEEM and is thus robust. In previous studies DWs with a width of $\SI{134}{nm}-\SI{184}{nm}$ \cite{weber2003magnetostrictive} up to several hundred nm \cite{arai2012three} were observed in bulk NiO crystals. In comparison to these values the DWs observed in our thin films are narrow. This indicates that the anisotropy in our NiO thin films stemming from the substrate-induced strain is larger compared to the bulk, favouring narrower DWs and in line with the observed change of the domain orientation from the $\{111\}$ planes in the bulk to the $[\pm5\ \pm5\ 19]$ orientations found here. In terms of applications as storage devices, such narrow domain walls are favourable, since the DW width is a limit for the storage density.

\section{\label{sec:conclusion}Conclusion}
By combining XMLD-PEEM imaging and current-induced switching we show that in insulating antiferromagnetic NiO/Pt thin films, circular shaped antiferromagnetic domain structures can be generated, indicating that the crystallographic restrictions present in bulk samples do not apply for thin films. We also show that the spins within the DW structure are always pointing in the same direction irrespective of their position in the circular path of the DW. This indicates that the domain walls are non-chiral, indicating the absence of strong Lifshitz invariants. Therefore, mechanisms predicting electrical switching of insulating antiferromagnetic films due to chirality dependent movement of DWs \cite{baldrati2018full, Gray2019} cannot explain the observed electical switching of the magnetic structure. Further, we show that the Néel vector across the DW rotates on the shortest path between the Néel vector orientation of the two adjacent domains thus, increasing the anisotropy energy density. However, such increment in anisotropy energy is compensated by the reduction of the DW width and the exchange energy density. Using micromagnetic simulations we show that such a behavior of the Néel vector can be explained due to strain distribution within the NiO thin film. Extracting the simulated DW profile and fitting it to the XMLD data of the DW, allows us to determine the width of the DW to $\Delta = 98 \pm 10$ nm. The observation of narrow DWs and their advantages for data storage devices underlines the importance of NiO in the context of AFM spintronics and the approach to make use of antiferromagnets as active elements in spintronic devices. For example, by varying the AFM film thickness, one could use the substrate-induced strain to tailor the DW width to optimize the storage density in devices relying on antiferromagnetic materials and by this make use of the advantages AFMs posses over FMs.

\begin{acknowledgments}
L.B. acknowledges the European Union’s Horizon 2020 research and innovation program under the
Marie Skłodowska-Curie Grant Agreements ARTES No.
793159. L.B., M.F., and M.K. acknowledge
support from the Graduate School of Excellence Materials
Science in Mainz (MAINZ) DFG 266, the DAAD (Spintronics network, Project No. 57334897). We acknowledge
that this work is funded by the Deutsche Forschungsgemeinschaft (DFG, German Research Foundation)—TRR
173-268565370 (Projects A01 and B02). This project has received funding from
the European Union’s Horizon 2020 research and innovation programme under Grant Agreement No. 863155
(s-Nebula). We acknowledge Diamond Light Source for
time on beamline I06 under proposal MM22448. This work is also supported
by ERATO “Spin Quantum Rectification Project” (Grant
No. JPMJER1402) and the Grant-in-Aid for Scientific
Research on Innovative Area, “Nano Spin Conversion Science” (Grant No. JP26103005), Grant-in-Aid for Scientific Research (S) (Grant No. JP19H05600), Grant-in-Aid
for Scientific Research (C) (Grant No. JP20K05297)
from JSPS KAKENHI. R.R. acknowledges support from the European Commission through the project 734187-SPICOLOST (H2020-MSCA-RISE-2016), the European Union's Horizon 2020 research and innovation program through the MSCA grant agreement SPEC-894006, Grant RYC 2019-026915-I funded by the MCIN/AEI/ 10.13039/501100011033 and by "ESF investing in your future", the Xunta de Galicia (ED431B 2021/013, Centro Singular de Investigación de Galicia Accreditation 2019-2022, ED431G 2019/03) and the European Union (European Regional Development Fund - ERDF). L.S.T. and G.F. acknowledge financial support
from the program “ASSEGNI DI RICERCA 2019” at the
University of Messina, the project PRIN 2020LWPKH7 funded by the Italian Ministry of University and Research, and PETASPIN association (www.petaspin.com).
\end{acknowledgments}

\appendix

\section{Micromagnetic simulations}
\label{AppendA}

The AFM state of the sample can be described by two sublattice magnetizations strongly coupled by the exchange interaction. For each of the sublattices the magnetization \textbf{m} follows the Landau-Lifshitz-Gilbert equation \cite{gomonay2010spin, puliafito2019micromagnetic, sanchez2020dynamics}

\begin{align}
\label{eq:S1}
\begin{cases}
    \frac{d\mathbf{m}_1}{dt} &= -\gamma_0 \mathbf{m}_1 \times \mathbf{H}_{\mathrm{eff},1} + \alpha \mathbf{m}_1 \times \frac{\mathbf{m}_1}{dt}\\
    \frac{d\mathbf{m}_2}{dt} &= -\gamma_0 \mathbf{m}_2 \times \mathbf{H}_{\mathrm{eff},2} + \alpha \mathbf{m}_2 \times \frac{\mathbf{m}_2}{dt}
\end{cases}
\end{align}

where $\gamma_0$ is the gyromagnetic ratio, $\alpha$ the Gilbert damping coefficient and

\begin{align}
\label{eq:S2}
    \mathbf{H}_{\mathrm{eff},i} = -\frac{1}{\mu_0 M_\mathrm{s}}\frac{\delta\epsilon}{\delta \mathbf{m}_i}
\end{align}

the effective field for the i-th sublattice, $M_s$ the sublattice saturation magnetization, $\mu_0$ the vacuum magnetic permeability and $\epsilon$ the energy density of the system. The energy density includes contributions from the exchange fields, the anisotropy field and magnetoelastic interaction. The expression for the exchange fields can be found in previous works \cite{puliafito2019micromagnetic, sanchez2020dynamics} while the bulk anisotropy energy and can be found in our previous work \cite{Schmitt2021}

\begin{align}
\label{eq:S3}
    \epsilon_{\alpha} (\mathbf{m}_{\alpha}) = K_{\alpha} (1-a_p(\mathbf{m}_{\alpha}))^2,
\end{align}

\begin{widetext}
    \begin{align}
    \label{eq:S4}
        a_1(\mathbf{m}) &= \left(1-\frac{1}{6}(m_x + m_y -2m_z)^2\right) \left(1-\frac{1}{6}(m_x - 2m_y +m_z)^2\right)\left(1-\frac{1}{6}(2m_x - m_y -m_z)^2\right) \nonumber \\
        a_2(\mathbf{m}) &= \left(1-\frac{1}{6}(2m_x - m_y +m_z)^2\right) \left(1-\frac{1}{6}(m_x + m_y + 2m_z)^2\right)\left(1-\frac{1}{6}(m_x - 2m_y -m_z)^2\right) \nonumber \\
        a_3(\mathbf{m}) &= \left(1-\frac{1}{6}(m_x - m_y -2m_z)^2\right) \left(1-\frac{1}{6}(2m_x + m_y -m_z)^2\right)\left(1-\frac{1}{6}(m_x + 2m_y +m_z)^2\right) \nonumber \\
       a_4(\mathbf{m}) &= \left(1-\frac{1}{6}(m_x - m_y +2m_z)^2\right) \left(1-\frac{1}{6}(2m_x + m_y +m_z)^2\right)\left(1-\frac{1}{6}(m_x + 2m_y -m_z)^2\right)
    \end{align}
\end{widetext}

Where $a_p$ allows us to choose between the four possible easy planes. The magnetoelastic contribution reads \cite{Schmitt2021, popov2020voltage} 

\begin{align}
\label{eq:S5}
    \epsilon_{\mathrm{me}} = K_{ij}^{\mathrm{ME}} m_{\alpha, i} m_{\alpha, j},
\end{align}
\begin{align}
\label{eq:S6}
    K_{ij}^{\mathrm{ME}} = b_{ijkl}e_{kl} = (b_0 \delta_{ij}\delta_{kl}+ b_1(\delta_{ik}\delta_{jl}+\delta_{ik}\delta_{jl}))e_{kl}.
\end{align}

$b_0$, $b_1$ are the components of the magnetoelastic tensor for NiO, and $e_{kl}$ is the strain tensor. It is possible to stabilize the equilibrium configurations that we observe in our sample assuming the off-diagonal elements are zero if we consider together the bulk anisotropy described by Eq. \ref{eq:S3} as it was done in our previous work \cite{Schmitt2021}. However, this anisotropy must comply with certain crystallographic restrictions. Although we do not consider the bulk anisotropy, it is possible to reproduce the Néel vector direction in the domains by consider the off-diagonal elements of the strain tensor. In that way, each domain have the same diagonal strains, the one measured $e_{11}=e_{22}=8.6\times 10^{-3}$ and $e_{33}=-7.1\times 10^{-4}$, but different off diagonal strains, $e_{12}=1.9\times 10^{-8}$, $e_{13}= \pm 4.6\times10^{-3}$ and $e_{23} =\pm 4.6\times 10^{-3}$. The other parameters considered for the simulations are: $A_{11} =\SI{5}{pJ/m}$, $A_0 = \SI{-5}{pJ/m}$, $M_s= \SI{0.35}{MA/m}$, $K_a = \SI{0.25}{MJ/m^3}$. It should be noticed that the magnetoelastic coupling coefficients must be set equal to $b_0+2b_1 = \SI{5.e7}{Jm^{-3}}$ if we consider the bulk anisotropy but $b_0+2b_1= \SI{5.e6} {Jm^{-3}}$ if we consider only the magnetoelastic contributions.

\section{Resolution determination}
\label{AppendB}

\begin{figure}[h!]
 \includegraphics[width=9cm]{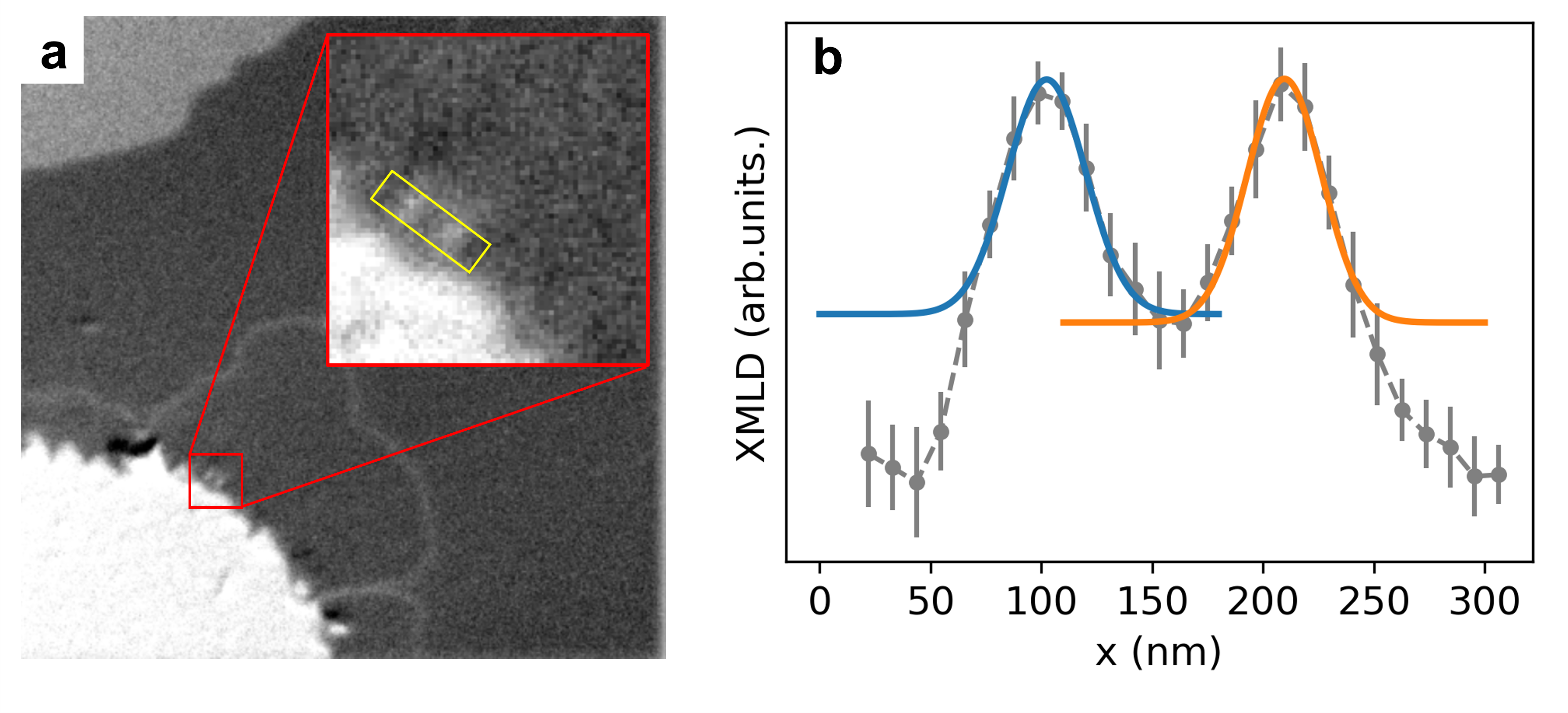}
 \caption{(a) XMLD-PEEM image taken with a field of view of $\SI{5.6}{\micro m}$ for $\gamma =$ 45° and $\omega =$ 75°. The inset shows a magnification of two small defects. Their line profile was taken in (b) to determine the resolution limit of the PEEM setup in this experiment by fitting Gaussian functions to the peaks.}
 \label{fig:resolution}
\end{figure}

The representation of structures in XMLD-PEEM images is limited by the spatial resolution of the microscope. The resolution influences how large structures appear in the images, or whether they can be resolved at all. Therefore, when determining the DW width, the minimum resolution needs to be determined first. This can be estimated using the line profile of the smallest visible defect. On the MgO(001)//NiO(10 nm)/Pt(2 nm) sample, the patterning edge contains small defects, as the XMLD-PEEM image in Fig. \ref{fig:resolution} (a) shows. The smallest defects that can be found are shown in the inset of Fig. \ref{fig:resolution} (a). A line profile along these two defects yields the intensity distribution shown in Fig. \ref{fig:resolution} (b). The error bars result from the standard deviation when averaging over five neighbouring pixels. The size of the defects is estimated by fitting a Gaussian function to the peaks in the XMLD signal. This yields a full width at half maximum (FWHM) of $\Delta_1 = 41 \pm \SI{4}{nm}$ for the first peak located at $\SI{102}{nm}$ and $\Delta_2 = 40 \pm \SI{4}{nm}$ for the peak located at $\SI{210}{nm}$. These values set an upper limit to the resolution minimum. This limited resolution can be taken into account when determining the DW width by fitting the XMLD signal convoluted with a Gaussian function with FWHM corresponding to the resolution limit \cite{boulle2016room}.

\bibliography{main}

\end{document}